\documentclass{IEEEcsmag}
\usepackage{flushend}
\usepackage[T1]{fontenc}
\usepackage[utf8]{inputenc}
\usepackage{dblfloatfix}
\usepackage{mathptmx}
\usepackage{soulutf8}
\usepackage[kerning,spacing]{microtype}
\usepackage{setspace}
\usepackage{graphicx}
\usepackage{adjustbox}
\usepackage{subcaption}
\usepackage{comment}
\usepackage{algpseudocode,algorithm}
\usepackage{pifont}
\usepackage{array}
\usepackage[table,xcdraw]{xcolor}
\usepackage{tabularx}
\usepackage{multirow}
\usepackage{makecell}
\usepackage{color}
\usepackage{nameref}
\usepackage[hyphens]{url}
\usepackage{booktabs}
\usepackage{colortbl}
\usepackage[sort]{natbib}
\usepackage[colorlinks,urlcolor=blue,linkcolor=blue,citecolor=blue]{hyperref}

\usepackage[framemethod=tikz]{mdframed}
\newcommand*{\circleNum}[1]{\tikz[baseline=(char.base)]{
              \node[shape=circle,fill,inner sep=0.9pt] (char) {\textcolor{white}{\scriptsize{#1}}};}}

\usepackage{kotex}

\usepackage{natbib}
\setcitestyle{numbers}

\jvol{XX}
\jnum{XX}
\paper{8}
\jmonth{}
\jname{IEEE Micro}
\pubyear{}

\newcommand\extrafootertext[1]{%
    \bgroup
    \renewcommand\thefootnote{\fnsymbol{footnote}}%
    \renewcommand\thempfootnote{\fnsymbol{mpfootnote}}%
    \footnotetext[0]{#1}%
    \egroup
}

\begin{document}

\title{Failure Tolerant Training with Persistent Memory Disaggregation over CXL}

\author{Miryeong Kwon, Junhyeok Jang, Hanjin Choi}
\affil{KAIST}

\author{Sangwon Lee, Myoungsoo Jung}
\affil{KAIST and Panmnesia}

\footernote{Appears in the IEEE Micro, Special Issue on Emerging System Interconnects, 2023 (Preprint)}

\begin{abstract}
This paper proposes \textsc{TrainingCXL} that can efficiently process large-scale recommendation datasets in the pool of disaggregated memory while making training fault tolerant with low overhead. To this end, i) we integrate persistent memory (PMEM) and GPU into a cache-coherent domain as Type-2. Enabling CXL allows PMEM to be directly placed in GPU's memory hierarchy, such that GPU can access PMEM without software intervention. \textsc{TrainingCXL} introduces computing and checkpointing logic near the CXL controller, thereby training data and managing persistency in an active manner. Considering PMEM's vulnerability, ii) we utilize the unique characteristics of recommendation models and take the checkpointing overhead off the critical path of their training. Lastly, iii) \textsc{TrainingCXL} employs an advanced checkpointing technique that relaxes the updating sequence of model parameters and embeddings across training batches. The evaluation shows that \textsc{TrainingCXL} achieves 5.2$\times$ training performance improvement and 76\% energy savings, compared to the modern PMEM-based recommendation systems.
\end{abstract}

\maketitle

\extrafootertext{\par\noindent\rule{2cm}{0.4pt} \\ This paper has been accepted at the IEEE Micro, Special Issue on Emerging System Interconnects on Jan 2023. The final version of the manuscript will be available soon and this material is presented to ensure the timely dissemination of scholarly and technical work.}

\chapterinitial{Deep learning} based recommendation systems take the majority of machine resources in diverse production servers and datacenters \cite{hazelwood2018applied}.
In practice, many production-level \textit{recommendation models} (RMs) require highly-accurate services to prevent Hyperscalers from undesirable losses in revenues.
This insists on large-sized models and feature vectors to train (i.e., embeddings), which are even much bigger than the largest deep neural network-based model such as Transformers.
For example, several studies have reported that the production-level RMs often consume more than tens of terabyte/petabyte memory spaces \cite{mudigere2022software}.

In addition, it is important for the RMs to be failure tolerant as they should be trained many days or weeks without an accuracy degradation.
To this end, the RMs periodically store their current training snapshots in persistent storage as checkpoints \cite{eisenman2022check}.
Even though the checkpoints are essential for system failure recovery, it is often considered the performance bottleneck in diverse computing domains, including the RMs \cite{eisenman2022check}. 

To address these challenges, several approaches have been proposed to employ high-performance solid-state drives (SSDs) and expand their host memory using SSDs as backend storage media \cite{wang2022merlin, zhao2019aibox}.
For example, \cite{wang2022merlin} store large-scale embedding tables while keeping only the feature vectors,
frequently accessed from the host computing-resources (CPU/GPU), in their local memory.
While this SSD-integrated memory expansion technique can handle the large-sized input data, they unfortunately suffer from severe performance degradation.
This is because RM's embedding lookup tasks often generate small-sized reads with a random pattern whereas SSDs are optimized for other types of bulk I/O operations.

Furthermore, all the prior approaches require explicit checkpoints for fault recovery.
Since embedding updates on SSDs can degrade the training performance significantly, the existing RMs utilize the SSDs for only memory expansion purposes.
Note that the write latency of SSDs is longer than the latency of all conventional memory operations by many orders of magnitude.
The writes also introduce many internal tasks, such as garbage collection, making the training performance unacceptable in many cases \cite{wu2022joint}.

We propose \textsc{TrainingCXL} that can efficiently process large-scale RM datasets in the underlying memory pool, disaggregated over \emph{compute express link} (CXL).
\textsc{TrainingCXL} makes deep learning training fault tolerant without imposing the checkpointing overhead as well. Our contributions can be summarized as follows:

\noindent $\bullet$ \textbf{Intelligent CXL memory expansion.}
\textsc{TrainingCXL} forms a non-volatile memory expander (\emph{CXL-MEM}) having many persistent memory modules (PMEM) as a Type-2 of CXL 3.0 \cite{cxl30}. We also introduce Type-2's \textit{device coherent agent} (DCOH) to GPU, called \emph{CXL-GPU}.
Since \textsc{TrainingCXL} integrates CXL-MEM and CXL-GPU into the same cache coherent domain, all the input/output embeddings are exchanged between those two without any software intervention running on the host CPU.
In our architecture, CXL-MEM employs simple computing and checkpointing logic along with a Type-2 endpoint controller, which can perform embedding operations and failure tolerance management near PMEM.

\noindent $\bullet$ \textbf{Batch-aware checkpoint.}
To achieve high training bandwidth, \textsc{TrainingCXL} lets CXL-GPU and CXL-MEM perform multi-layer perceptrons (MLP) and embedding operations simultaneously.
However, the model and embedding updates for the end of each batch process make the training latency yet longer.
To address this, we propose \emph{batch-aware checkpoint} that is aware of each shape of the individual batch and performs undo logging in background.
Practically, such a background undo logging scheme is infeasible for most applications as the target location where the system needs to update is unavailable before their computation completes.
Since we can know the next batch's embedding structure at each training stage in advance, our scheme logs appropriate models and embeddings to CXL-MEM in parallel with RM's training tasks.

\noindent $\bullet$ \textbf{Embedding lookup and checkpoint relaxation.}
While our batch-aware checkpoint can hide the relatively long latency of CXL-MEM's writes, the training performance can yet degrade owing to read-after-write (RAW) caused by embedding operations between two adjacent batches;
RMs read some of newly-updated embeddings at the beginning of each batch, but the reads can be stalled because of the writes (for the embedding updates) issued right before.
Unfortunately, the training also suffers performance degradation when our batch-aware checkpoint is slower than the MLP computation in CXL-GPU.
\textsc{TrainingCXL} relaxes the order of embedding lookups and checkpoints, and it reschedules them to remove the RAW conflict issues and checkpointing overhead.
This relaxation can make CXL-GPU and CXM-MEM operations fully overlap each other, thereby improving the training bandwidth further.

Our evaluation results show that training performance can be improved by 5.2$\times$. \textsc{TrainingCXL} also achieves 76\% energy savings, compared to modern PMEM-based recommendation systems.

\section{Background}
\label{sec:background}
\begin{figure}
    \centering
    \includegraphics[width=1\linewidth]{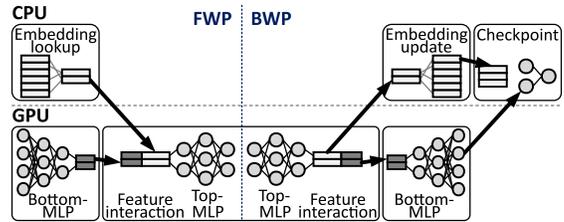}
    \vspace{-8pt}
    \caption{Training process of DLRM.} \label{fig:back_dlrm}
    \vspace{-8pt}
\end{figure}

\subsection{Recommendation Model Training}
Meta AI's DLRM \cite{naumov2019deep} is a representative RM used for personalized item recommendations.
To achieve a higher accuracy and better representation capacity, DLRM exploits both sparse features (categorical information) and dense features (continuous information).
Because of these distinct characteristics between the spare and dense features, they are encoded by different types of processing operations before going through the main training task, called \textit{Top-MLP}.
While the dense features are learned and encoded into input(s) of the top-MLP using conventional matrix-multiplication operations, called \textit{Bottom-MLP}, the sparse features are processed by a set of embedding operations (e.g., table lookup/update).

Figure \ref{fig:back_dlrm} shows a single-batch training process of DLRM.
Considering different levels of the computing intensiveness, the bottom-MLP operations are all performed in GPU, whereas the embedding operations are practically processed at the host-side (CPU).
For these embedding operations, the host reads the target embedding vectors from the underlying storage by referring to the corresponding table indices (residing in the sparse features).
Note that production-scale embedding tables often exceed tens to hundreds TBs, which unfortunately cannot be accommodated in the target system's local memory.
The host then aggregates the retrieved embedding vectors using simple arithmetic (e.g., add/subtract) and generates new embedding vectors as another input of the top-MLP.
As the bottom-MLP and embedding operations are simultaneously processed in different places (i.e., GPU and CPU), each encoded input data for the top-MLP can be prepared in parallel thereby saving the training time at some extent \cite{wang2022merlin}.
To put the encoded inputs into a same vector space, GPU performs a \textit{feature interaction} (e.g., concatenation) and the top-MLP, which trains the result of the feature interaction in this forward-propagation step (FWP).
The backward-propagation step of RM training (BWP) is similar to FWP, but all its operations are processed in the reverse order.
BWP takes gradient errors as an input (instead of FWP's input features) and updates model parameters/embeddings to improve the model accuracy.
The gradient errors are calculated by the differences between the FWP results and labels (truth grounds).
Note that all the updated model parameters (e.g., MLPs' weights and embeddings) should be checkpointed in the underlying storage for each end of all the batches.

\begin{figure}
    \centering
    \includegraphics[width=1\linewidth]{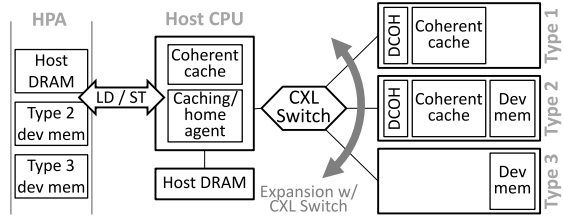}
    \vspace{-8pt}
    \caption{CXL architecture and CXL device types.} \label{fig:back_cxl}
    \vspace{-12pt}
\end{figure}

\begin{figure*}[b]
    \vspace{-4pt}
    \addtocounter{figure}{1}
    \centering
    \includegraphics[width=1\linewidth]{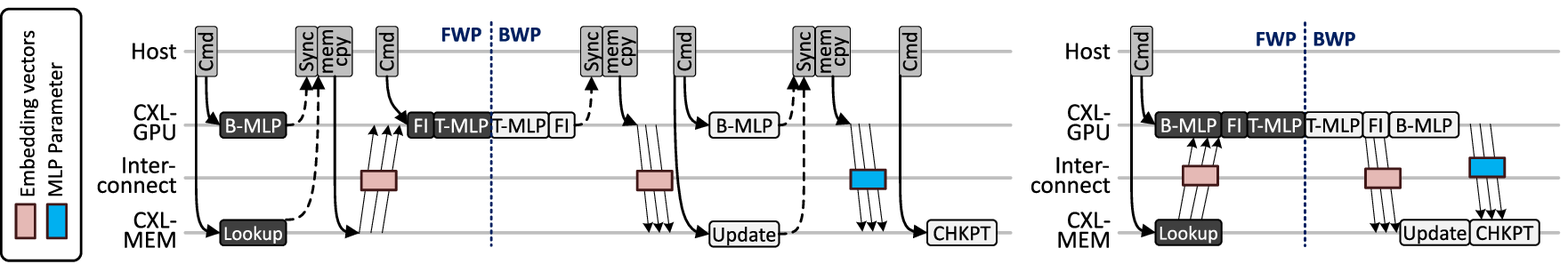}
    \begin{subfigure}{\linewidth}
      \centering
      \renewcommand*{\arraystretch}{0.3}
      \begin{tabularx}{\textwidth}{
        p{\dimexpr.7\linewidth-2\tabcolsep-1.3333\arrayrulewidth}
        p{\dimexpr.3\linewidth-2\tabcolsep-1.3333\arrayrulewidth}
      }
        \vspace{-8pt} \caption{Conventional software framework.} \label{fig:design_tensor1}
      & \vspace{-8pt} \caption{TrainingCXL's hardware.} \label{fig:design_tensor2}
      \end{tabularx}
    \end{subfigure}
    \vspace{-14pt}
    \caption{Comparison of recommendation model training procedure.} \label{fig:design_tensor}
\end{figure*}

\begin{figure}
  \addtocounter{figure}{-2}
    \centering
    \includegraphics[width=1\linewidth]{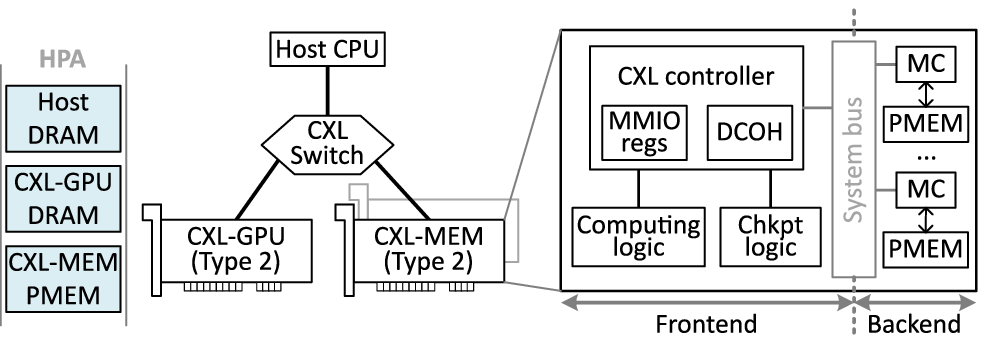}
    \begin{subfigure}{\linewidth}
      \centering
      \renewcommand*{\arraystretch}{0.3}
      \begin{tabularx}{\textwidth}{
        p{\dimexpr.50\linewidth-2\tabcolsep-1.3333\arrayrulewidth}
        p{\dimexpr.50\linewidth-2\tabcolsep-1.3333\arrayrulewidth}
      }
        \vspace{-4pt} \caption{Cache-coherent domain.} \label{fig:overview_hw1}
      & \vspace{-4pt} \caption{Internals of CXL-MEM.} \label{fig:overview_hw2}
      \end{tabularx}
    \end{subfigure}
    \vspace{-13pt}
    \caption{\textsc{TrainingCXL}'s system architecture.} \label{fig:overview_hw}
    \vspace{-15pt}
  \addtocounter{figure}{1}
\end{figure}

\subsection{Compute Express Link (CXL)}
CXL is an open industry standard interconnect, which allows multiple heterogeneous (computing) devices to share large-scale memory spaces in a cache-coherent manner.
Figure \ref{fig:back_cxl} shows a system architecture that enables CXL.
The system consists of three essential hardware components: i) CXL-enabled host CPU(s), ii) CXL switch(es), and iii) CXL device(s).
CXL devices can be incarnated by leveraging the design of conventional peripheral devices (e.g., accelerator and memory expander), but they should accommodate appropriate CXL protocol interfaces for their specific purposes.
CXL switches can interconnect the host CPU(s) and multiple CXL devices in a disaggregated manner. Note that a CXL network (per CPU's root-complex) can have upto 4095 CXL devices, which can all the host to secure sufficient memory space to use \cite{cxl30}.
All these CXL hardware components can have a unified memory space, referred to as \textit{Host Physical Address} (HPA) as their shared memory pool.
To this end, CXL supports three sub-protocols (\textit{CXL.io}, \textit{CXL.cache}, and \textit{CXL.mem}).

Based on how to combine these sub-protocols, CXL devices can be classified into Type-1, Type-2, and Type-3.
CXL.io is mandatory for all the types of hardware, allowing a CXL device to expose device registers to HPA as memory-mapped IO (MMIO) registers.
Using this protocol, the host CPU can discover or configure the underlying CXL devices (by reading/writing the MMIO registers).
In addition to CXL.io, Type-1 and Type-3 need to support CXL.cache and CXL.mem, respectively.
Note that Type-2 is recommended to have all three sub-protocols as it is designed toward having both computing and memory resources at the backend.
The memory resources of Type-2 is accessed by host through CXL.mem, while Type-2's computing resources can buffer data residing in the HPA into its internal cache with CXL.cache.
Internal cache's cacheline states are tracked by \textit{device coherency engine} (DCOH) \cite{cxl30} to guarantee cache-coherency with other CXL devices.

\section{Persistent Memory Disaggregation}

\label{sec:overview}
\textsc{TrainingCXL} disaggregates memory devices from CPU/GPU and integrates all of them into a single system over CXL.
Our CXL-enabled memory expander employs PMEMs, which exhibit similar performance to DRAM and provide large-scale memory capacity.
Thanks to the backend PMEM, \textsc{TrainingCXL} can reduce the overhead imposed by heavy reads/writes (for embeddings), compared to the existing RM system.
In addition, \textsc{TrainingCXL} uses PMEM for memory expansion as well as leverages its non-volatile characteristics to support failure tolerant training with low overhead.
Lastly, \textsc{TrainingCXL} alleviates the data movement overhead between GPU and memory expander by exploiting the advantages that CXL sub-protocols offer.

\subsection{System Architecture}
Figure \ref{fig:overview_hw1} shows an overview of the proposed \textsc{TrainingCXL}'s system architecture.
\textsc{TrainingCXL} proposes two CXL devices: a CXL-enabled GPU (\textit{CXL-GPU}) and a PMEM-based memory expander (\textit{CXL-MEM}).
These are designed as Type-2 and connected to a host CPU through CXL Switch(es).
Because of their device type, CXL-MEM's internal memory can be exposed to CXL-GPU's local memory and vice versa.
To accelerate RM processing, CXL-MEM manages all the embedding operations instead of the host CPU.
Since the host CPU is free from processing embeddings in \textsc{TrainingCXL}, it is only responsible for running RM training software (e.g., PyTorch or TensorFlow).

\noindent \textbf{Designing CXL-MEM.}
Figure \ref{fig:overview_hw2} illustrates CXL-MEM composed by the frontend and backend modules.
The backend employs multiple PMEMs and corresponding memory controllers to achieve a high degree of data parallelism for large-scale embedding tables as well as model checkpoints.
The aggregated memory space of PMEMs is exposed to CXL-MEM's frontend through system interconnect, which consists of i) a CXL controller that implements all the three CXL sub-protocols, ii) a computing logic that processes embedding operations (lookup/update), and iii) a checkpointing logic that automatically creates the model checkpoints.
To initialize computing logic, the host CPU sets CXL-MEM's MMIO registers with embedding vector length and learning rate, which are required by embedding operations.
In addition, it is important to store the MLP parameters' memory address and the size of MLP parameters to allow checkpointing logic read MLP parameters from CXL-GPU.
We will explain how this information is used by CXL-MEM's computing and checkpointing logic with more details, in the ``\nameref{subsec:chkpt}'' section.

\subsection{CXL-based Automatic Data Movement}
Since \textsc{TrainingCXL} adopts a heterogeneous computing system, it is inevitable to move RM training's intermediate data across a set of CXL devices.
However, this data movement has been managed by the RM training software running on the host CPU.
To remove non-negligible software overhead, \textsc{TrainingCXL} proposes a hardware automation approach that moves data by leveraging CXL hardware components.

\begin{figure}
  \centering
  \includegraphics[width=1\linewidth]{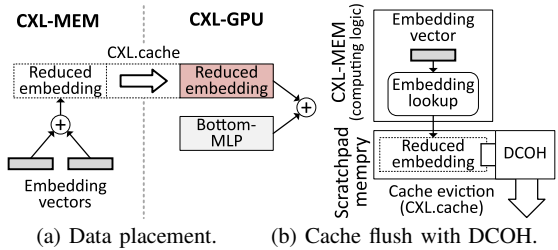}
  \begin{subfigure}{\linewidth}
    \centering
    \renewcommand*{\arraystretch}{0.3}
    \begin{tabularx}{\textwidth}{
      p{\dimexpr.45\linewidth-2\tabcolsep-1.3333\arrayrulewidth}
      p{\dimexpr.55\linewidth-2\tabcolsep-1.3333\arrayrulewidth}
      }
      \vspace{-3pt} \caption{Data placement.} \label{fig:design_dcoh1}
    & \vspace{-3pt} \caption{Cache flush with DCOH.} \label{fig:design_dcoh2}
  \end{tabularx}
  \end{subfigure}
  \vspace{-12pt}
  \caption{CXL-based automatic data movement.} \label{fig:design_dcoh}
  \vspace{-12pt}
\end{figure}

\noindent \textbf{Performance impacts.}
Figure \ref{fig:design_tensor1} demonstrates the performance impact of the conventional software-based data movement approaches.
Using the FWP as an example, the software offloads bottom-MLP and embedding lookup to CXL-GPU and CXL-MEM, respectively.
They notify the completion of the operation to the software through \texttt{cudaStreamSynchronize}; after that, the software can start to copy the new embedding vectors from CXP-MEM to CXL-GPU (e.g., \texttt{cudaMemcpy}) and request CXL-GPU to perform feature interaction and top-MLP.
These software overhead can be eliminated with CXL-based automatic data movement as shown in Figure \ref{fig:design_tensor2}.
\textsc{TrainingCXL}'s CXL hardware components are responsible for data movement and RM training operations can be synchronized by checking whether all input data are ready or not.

\noindent \textbf{Design of hardware automation approach.}
The insight behind CXL-based automatic data movement is to leverage CXL.cache to transfer data.
To move the data using CXL.cache, the data should be stored where it will be used as an operation input.
For example, the input data of feature interaction (e.g., reduced embedding vector) should be stored in the CXL-GPU's device memory and cached in the CXL-MEM's internal cache as shown in Figure \ref{fig:design_dcoh1}. Then the CXL-MEM's DCOH flushes every cacheline of the reduced embedding vector, which is updated by embedding lookup as shown in Figure \ref{fig:design_dcoh2}. Similarly, in the case of BWP, the input data of the embedding update (e.g., embedding gradient) should be stored in the CXL-MEM's device memory and cached in the CXL-GPU. Then the embedding gradient is flushed by the CXL-GPU's DCOH and transferred to the CXL-MEM.

\begin{figure}
  \centering
  \includegraphics[width=1\linewidth]{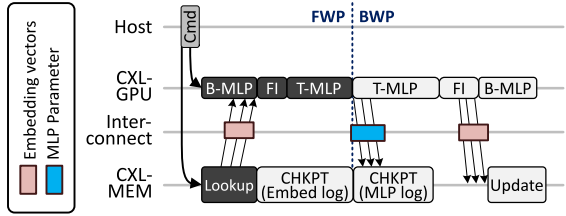}
  \vspace{-8pt}
  \caption{Execution of batch-aware checkpoint.}\label{fig:design_timeline}
  \vspace{-12pt}
\end{figure}

\section{Failure Tolerance Management}

\label{sec:design}
This section will explain 1) how the checkpointing overhead is removed from the critical path of RM training, and 2) how CXL-MEM's checkpointing logic can perform checkpointing automatically in the background.

\subsection{Batch-aware Checkpoint}
The conventional SSD-based failure tolerance management is performed in a redo log manner.
In other words, the updated embedding vectors and bottom/top-MLP parameters have been permanently stored at the end of each training epoch (before starting the next batch training).
To take checkpointing off the critical path of RM training, \textsc{TrainingCXL} proposes a batch-aware checkpoint that performs checkpointing in an undo log manner.
It leverages RM training's characteristic that embedding vector indices to be updated can be known in advance even if the batch training is not completed yet.
Since the sparse features include that information, RM training software sets them in the MMIO register for every batch to store embedding and MLP logs by utilizing the idle time of CXL-MEM as shown in Figure \ref{fig:design_timeline}.

\subsection{CXL-MEM's Checkpoint Support}\label{subsec:chkpt}
To support batch-aware checkpoint, we first split the CXL-MEM's memory space into data and log regions.
Each of these regions is for computing logic and checkpointing logic to store embedding tables and embedding/MLP logs, respectively.
Since embedding/MLP logs are managed in the separated region, CXL-MEM can easily recover the RM model by referring to the log region when there is a power failure.

\begin{figure}
  \centering
  \includegraphics[width=1\linewidth]{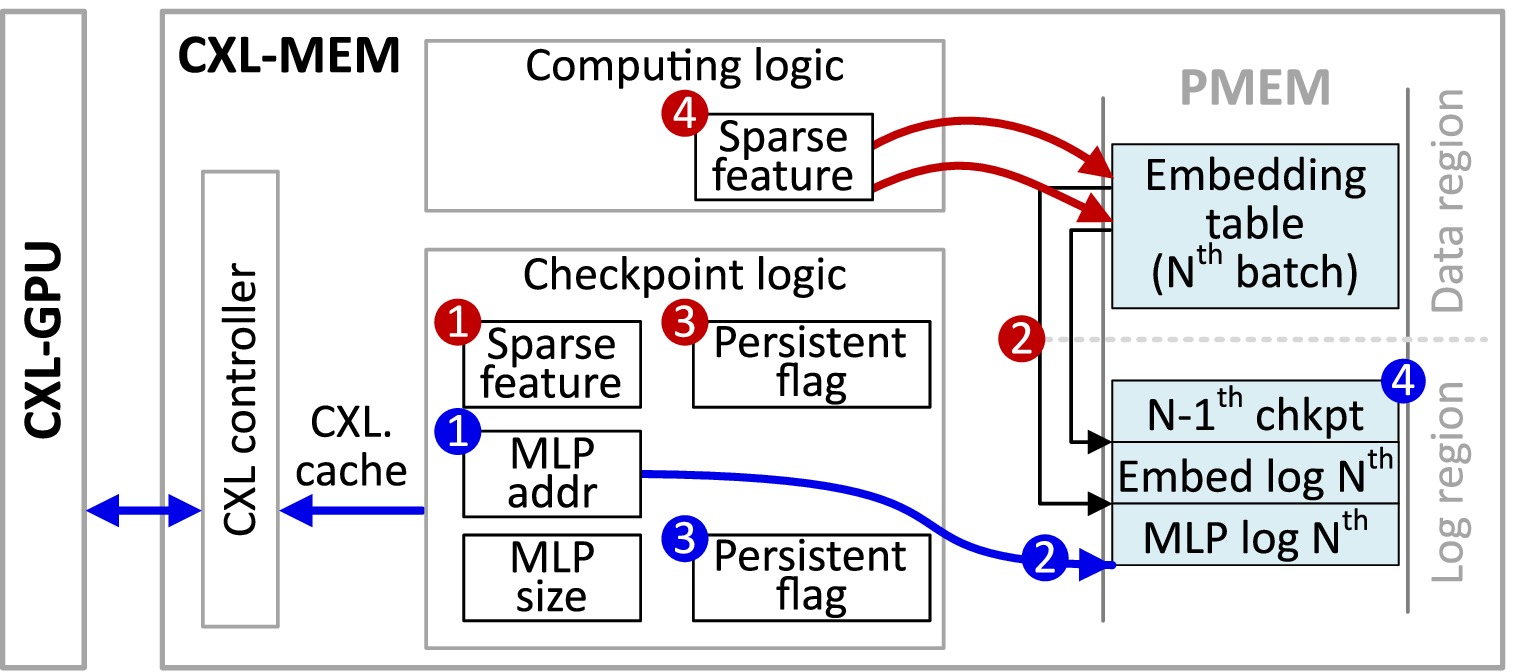}
  \vspace{-8pt}
  \caption{CXL-MEM's automatic checkpoint.}\label{fig:design_chkpt}
  \vspace{-12pt}
\end{figure}

\noindent \textbf{Embedding logging.}
Figure \ref{fig:design_chkpt} shows how the checkpointing logic can generate checkpoint for embedding table. The checkpointing logic refers to the embedding vector indices from the sparse feature (\textcolor{red}{\circleNum{1}}), and creates embedding log by copying the embedding vectors from the data region to the log region (\textcolor{red}{\circleNum{2}}).
When embedding log is permanently stored, checkpointing logic sets persistent flag as true (\textcolor{red}{\circleNum{3}}).
Since checkpointing logic already backed up the embedding vectors in the background, the embedding table in the data region can be directly updated during embedding update operation (\textcolor{red}{\circleNum{4}}).
Even if a power failure occurs during an embedding update, training can be resumed from that batch if the persistent flag is set.

\noindent \textbf{MLP logging.}
Unlike embedding logging, the bottom/top-MLP parameters are stored in the CXL-GPU.
Therefore, the checkpointing logic should send CXL.cache requests by referring to MLP parameters' memory address and size stored in the MMIO registers (\textcolor{blue}{\circleNum{1}}).
When the MLP parameters are transferred from CXL-GPU, CXL-MEM stores them in the log region (\textcolor{blue}{\circleNum{2}}).
By comparing the number of transferred MLP parameters and the MLP parameters' size stored in the MMIO register, checkpointing logic can recognize whether the MLP log is fully checkpointed. When all MLP parameters are safely stored, checkpointing logic sets the persistent flag as true for the MLP log (\textcolor{blue}{\circleNum{3}}).
If the persistent flag is set for both the embedding and MLP log, checkpointing logic deletes the old checkpoint written in the previous batch (\textcolor{blue}{\circleNum{4}}).

\section{Relaxation of Failure Tolerant Training}

\label{sec:opt}
\textsc{TrainingCXL} proposes training relaxation techniques that can eliminate performance degradation caused by PMEM-based CXL-MEM design.

\subsection{Relaxed Embedding Lookup}
PMEM's read performance is similar to that of DRAM. However, it can be degraded if a read is requested right after a write for the same physical layout of PMEM.
This phenomenon is known as read-after-write (RAW) \cite{park2018bibim}, and can be found between the embedding update of the N\textsuperscript{th} batch and the embedding lookup of the (N+1)\textsuperscript{th} batch. \cite{kwon2022training} analyzed that 80\% of embedding vectors are trained for the consecutive batches, and this report indicates that RAW can be frequently observed during RM training.

\begin{figure}
  \centering
  \includegraphics[width=1\linewidth]{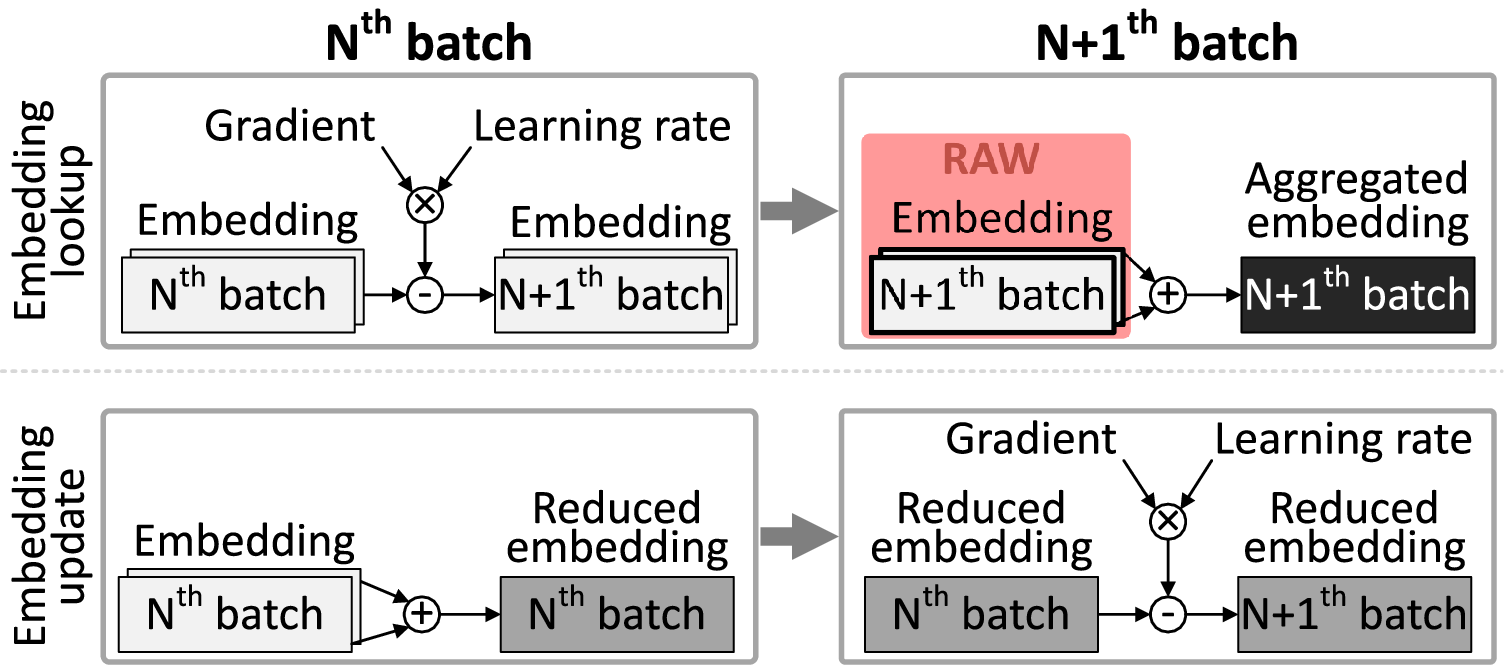}
  \vspace{-8pt}
  \caption{Dependency (top) vs. relaxed (bottom).} \label{fig:opt_embed}
  \vspace{-12pt}
\end{figure}

\noindent \textbf{Relaxation.}
As shown in the top of Figure \ref{fig:opt_embed}, (N+1)\textsuperscript{th} batch's embedding lookup has been performed when (N+1)\textsuperscript{th} batch's embedding table is prepared (by completing the N\textsuperscript{th} batch). In other words, there is operation dependency between the N\textsuperscript{th} batch's embedding update and the (N+1)\textsuperscript{th} batch's embedding lookup. However, since the embedding lookup and update are composed of addition/subtraction-based arithmetic, the operation dependency between embedding lookup and update can be relaxed by utilizing the commutative property of addition. Thus, \textsc{TrainingCXL} proposes relaxed embedding lookup as shown in the bottom of Figure \ref{fig:opt_embed}.
The conventional (N+1)\textsuperscript{th} batch's embedding lookup is now split into two steps.
First, the embedding lookup is performed at N\textsuperscript{th} batch with the N\textsuperscript{th} batch's embedding table.
After that, the embedding update for the reduced embedding vector is performed when the embedding gradient is ready.

\subsection{Relaxed Batch-aware Checkpoint}
Although batch-aware checkpoint can mitigate the slow PMEM writes by overlapping checkpointing with CXL-GPU's feature interaction and top-MLP, checkpointing overhead can be observed if the checkpoint time is longer than the idle time that CXL-MEM can exploit.

\begin{figure}
  \centering
  \begin{subfigure}[]{.39\linewidth}
    \includegraphics[width=1\linewidth]{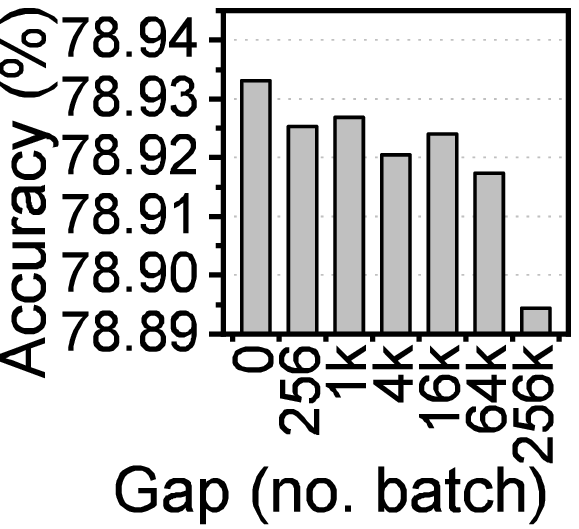}
  \end{subfigure}
  \begin{subfigure}[]{.55\linewidth}
    \includegraphics[width=1\linewidth]{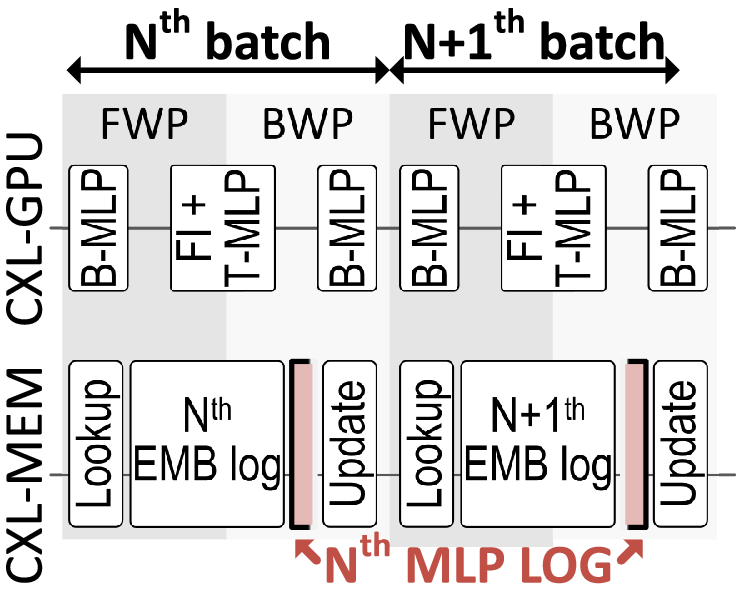}
  \end{subfigure}
  \begin{subfigure}{\linewidth}
    \centering
    \renewcommand*{\arraystretch}{0.3}
    \begin{tabularx}{\textwidth}{
      p{\dimexpr.4\linewidth-2\tabcolsep-1.3333\arrayrulewidth}
      p{\dimexpr.6\linewidth-2\tabcolsep-1.3333\arrayrulewidth}
    }
      \vspace{-9pt} \caption{Accuracy.} \label{fig:opt_chkpt1}
    & \vspace{-9pt} \caption{Schedule.} \label{fig:opt_chkpt2}
    \end{tabularx}
  \end{subfigure}
  \vspace{-18pt}
  \caption{Relaxed batch-aware checkpoint.} \label{fig:opt_chkpt}
  \vspace{-12pt}
\end{figure}

\noindent \textbf{Relaxation.}
Figure \ref{fig:opt_chkpt1} shows the training accuracy according to the batch number difference between the embedding and the MLP log.
As shown in Figure, the accuracy degradation satisfies the business needs (0.01\%) \cite{eisenman2022check} even when the batch gap of the two logs differs by hundreds.
This observation indicates that the bottom/top-MLP does not need to be checkpointed for every batch, although the embedding log should be permanently stored for every batch since the original embedding tables are updated for every batch. Thus, \textsc{TrainingCXL} proposes a relaxed batch-aware checkpoint that can schedule MLP logging across multiple batches as shown in Figure \ref{fig:opt_chkpt2}. Since the purpose of checkpointing relaxation is to hide its overhead from user experience, the MLP logging should be stopped when CXL-GPU completes the top-MLP operation. However, it is difficult for CXL-MEM to know whether CXL-GPU completes the MLP operation or not.
Thus, CXL-GPU supports this checkpointing relaxation by responding to the CXL-MEM's CXL.cache request only when it processes feature interaction and top-MLP.

\section{Evaluation}

\label{sec:evaluation}
\begin{figure}[b]
	\vspace{-8pt}
	\centering
    \includegraphics[width=.9\linewidth]{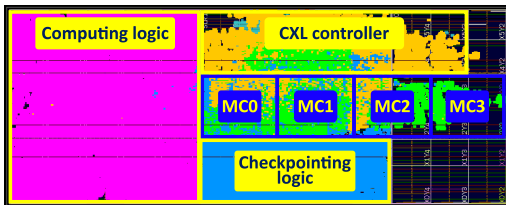}
	\vspace{-4pt}
	\caption{Prototype of CXL-MEM.} \label{fig:floor_plan}
\end{figure}

\noindent \textbf{Hardware prototype.}
The CXL-enabled RISC-V host CPU, CXL switch, CXL-GPU, and CXL-MEM are prototyped using a set of Xilinx Alveo U200s and a customized FPGA hardware devices.
Figure \ref{fig:floor_plan} shows CXL-MEM's hardware prototype as an example, which  includes a CXL device (3.0) controller, four memory controllers, computing/checkpointing logic. The computing logic consists of a set of adders, multipliers, and scratchpad memory to store interim embedding vectors.
The checkpointing logic has a CXL DMA engine and two counters, which deal with data copying for the embedding and MLP log. Note that the memory controllers emulate the PRAM latency by delaying DRAM responsiveness to make its memory performance similar to PMEM \cite{wang2020characterizing}. For CXL-GPU, we port Vortex \cite{tine2021vortex} into our platform using Xilinx IPs (rather than its original Altera IPs). However, its instruction set architecture cannot support diverse CUDA kernels that our RM systems and workloads use. We thus emulate the kernel latency in Vortex by replaying per-batch MLP computation cycles, which are extracted from 3090 GPU. The software interfaces such as \texttt{cudaMemcpy} are implemented using Vortex's DMA engine working on a PCIe address space that our CXL and Xilinx IPs handle.

\begin{table}
    \setlength{\tabcolsep}{3.5pt}
    \renewcommand{\arraystretch}{1.2}
    \centering
    \resizebox{\linewidth}{!}{
        \begin{tabular}{ccccc}
            \toprule
            \textbf{}      & \textbf{\begin{tabular}[c]{@{}c@{}}Embedding table storage\end{tabular}}
                           & \textbf{CPU}
                           & \textbf{Memory}
                           & \textbf{GPU} \\ \midrule
            \texttt{SSD}   & Intel 750 800GB
                           & \multirow{6}{*}{\begin{tabular}[c]{@{}c@{}}Intel\\ i5-9600K\\ 3.7GHz\end{tabular}}
                           & \multirow{6}{*}{\begin{tabular}[c]{@{}c@{}}4$\times$16GB\\ DRAM\end{tabular}}
                           & \multirow{6}{*}{\begin{tabular}[c]{@{}c@{}}NVIDIA\\ RTX 3090 \\ (Emulated by Vortex)\end{tabular}} \\ \cmidrule{1-2}
            \texttt{PMEM}  & Intel Optane PMEM 512GB                                                                           & & & \\ \cmidrule{1-2}
            \texttt{PCIe}  & \multirow{4}{*}{\begin{tabular}[c]{@{}c@{}}Xilinx Alveo U200 \& \\ Emulating PMEM 64GB\end{tabular}} & & & \\
            \texttt{CXL-D} &                                                                                                   & & & \\
            \texttt{CXL-B} &                                                                                                   & & & \\
            \texttt{CXL}   &                                                                                                   & & & \\ \bottomrule
        \end{tabular}
    }
    \caption{Test environment.} \label{tab:dev_info}
    \vspace{-5pt}
\end{table}

\begin{table}[b]
    \vspace{-10pt}
    \setlength{\tabcolsep}{3.5pt}
    \centering
    \resizebox{\linewidth}{!}{
        \begin{tabular}{|c|cc|cc|}
            \hline
            \multirow{2}{*}{} & \multicolumn{2}{c|}{Latency (vs. DRAM)} & \multicolumn{2}{c|}{Bandwidth (vs. DRAM)} \\ \cline{2-5}
                              & \multicolumn{1}{c|}{Read}    & Write    & \multicolumn{1}{c|}{Read}     & Write     \\ \hline
            PMEM              & \multicolumn{1}{c|}{3x}      & 7x       & \multicolumn{1}{c|}{0.6x}     & 0.1x      \\ \hline
            SSD               & \multicolumn{2}{c|}{165x}               & \multicolumn{2}{c|}{0.02x}                \\ \hline
            \end{tabular}
    }
    \caption{Device performance characteristics (normalized to DRAM performance).} \label{tab:dev_assum}
\end{table}

\noindent \textbf{Test configurations.}
We prepare three different configurations, SSD (\texttt{SSD}), PMEM (\texttt{PMEM}), and PCIe-attached PMEM (\texttt{PCIe}) based on the underlying media where the embedding tables are stored into. While embedding operations of \texttt{SSD} and \texttt{PMEM} are performed on the host CPU, \texttt{PCIe} is capable of near-data processing like our CXL-MEM. Moreover, \texttt{SSD} leverages host DRAM to cache embedding vectors. We also prepare three \textsc{TrainingCXL} configurations to breakdown our contributions: 1) CXL-MEM without any scheduling supports (\texttt{CXL-D}), 2) \texttt{CXL-D} with batch-aware checkpoint (\texttt{CXL-B}), and 3) \texttt{CXL-B} with relaxed training techniques (\texttt{CXL}).
Table \ref{tab:dev_info} summarizes the detailed hardware specifications we used for all test configurations, while Table \ref{tab:dev_assum} lists up device performance characteristics such as latency and bandwidth for read or write (which are normalized to that of DRAM).

\noindent \textbf{Model configuration.}
We use open source DLRM benchmark and prepare four recommendation system models (RM) for evaluation as listed in Table \ref{tab:rm_info}.
RM1, RM2, and RM3 are based on model parameters from \cite{gupta2020deeprecsys}, while RM4 is based on \cite{naumov2019deep}.
RM1 and RM2 are embedding-intensive models, requiring the lookup of 80 embedding vectors per embedding table.
In particular, RM2 has 4$\times$ many embedding tables than RM1, making it the most embedding-intensive model among RM1$\sim$4.
On the other hand, RM3 and RM4 are MLP-intensive models because the number of embedding vectors to lookup or the number of embedding tables is not that many compared to RM1 and RM2.
Note that we set the embedding table larger and the bottom-MLP deeper than the original model to reflect increases in dense and sparse features in the real world.
In addition, we consider Criteo Kaggle's embedding table access distribution when randomly generating sparse feature input for RM1$\sim$3 to evaluate the RAW impact similar to the real datasets.

\begin{table}
    \centering
    \setlength{\tabcolsep}{3.5pt}
    \renewcommand{\arraystretch}{1}
    \resizebox{\linewidth}{!}{
        \begin{tabular}{lcccc}
            \toprule
                                & \textbf{RM1}  & \textbf{RM2} & \textbf{RM3} & \textbf{RM4}  \\ \hline
            Input data set      & Random        & Random       & Random       & Criteo Kaggle \\
            \rowcolor[HTML]{EFEFEF}
            Features dim        &            32 &           32 &           32 &   		   16 \\
            \# Dense features   &            13 &           13 &           13 &            13 \\
            \rowcolor[HTML]{EFEFEF}
            \# Embed. table     &            20 &  			80 &           20 &   		   52 \\
            \# Sparse features  &            80 &           80 &  		   20 &    			1 \\

            \rowcolor[HTML]{EFEFEF}
            Bottom-MLP          & \begin{tabular}[c]{@{}c@{}}13-8192-\\2048-32\end{tabular}
                                & \begin{tabular}[c]{@{}c@{}}13-8192-\\2048-323\end{tabular}
                                & \begin{tabular}[c]{@{}c@{}}13-10240-\\4096-32\end{tabular}
                                & \begin{tabular}[c]{@{}c@{}}13-16384-\\2048-512-16\end{tabular}    \\
            Top-MLP             & 256-64-1      & 512-128-1
                                                              & 512-128-1
                                                                              & 512-128-1           \\ \bottomrule
        \end{tabular}
    }
    \caption{Recommendation system models.} \label{tab:rm_info}
    \vspace{-12pt}
\end{table}

\subsection{Overall Training Latency}
\begin{figure}
	\centering
	\includegraphics[width=1\linewidth]{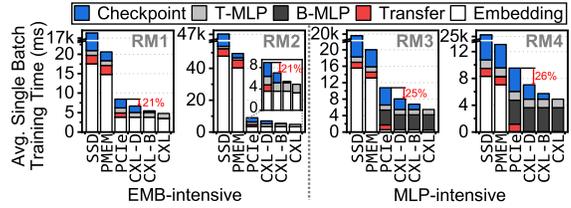}
	\vspace{-8pt}
	\caption{Training time breakdown.} \label{fig:rm_overall}
	\vspace{-10pt}
\end{figure}

Figure \ref{fig:rm_overall} shows RMs' average batch training time.
For the embedding-intensive models (RM1/RM2), \texttt{PMEM} exhibits 949$\times$ faster RM training time (including \texttt{T-MLP}, \texttt{B-MLP}, \texttt{Transfer}, and  \texttt{Embedding}, except for \texttt{Checkpoint}) than \texttt{SSD}, on average.
This is because PMEM can be accessed in a byte-addressable manner and its read/write is faster than SSD.
Since embedding tables are stored across multiple PMEMs, \texttt{PCIe}, \texttt{CXL-D}, \texttt{CXL-B}, and \texttt{CXL} can parallelize the embedding vector accesses as well as increase the performance of embedding operations.
Unfortunately, acceleration of embedding operations with NDP-capable PMEM does not work well for the MLP-intensive models compared to embedding-intensive models.
This is because MLP-intensive models exhibit long latency of \texttt{B-MLP} than embedding-intensive models, thereby failing to fully overlap that latency with \texttt{Embedding}.
Specifically, \texttt{CXL-D} shows 23\% training time reduction compared to \texttt{PCIe} on average.
Its performance benefits come from two reasons. First, all software overheads caused by \texttt{cudaStreamSynchronize} and \texttt{cudaMemcpy} are eliminated by automatic data movement. Second, CXL-MEM's checkpointing logic can directly examine the MLP parameters updated by CXL-GPU (for each batch) through CXL.cache, which can hide the overhead behind the time to compute (embedding operations). \texttt{CXL-B} further improves training performance than \texttt{CXL-D} by overlapping the checkpointing with the CXL-GPU operations.
Finally, our \texttt{CXL} can reduce training time by 14\% compared to \texttt{CXL-B} by eliminating checkpointing and RAW overheads through relaxed training techniques.

\subsection{Resource Utilization Analysis}

\begin{figure}
    \centering
    \begin{subfigure}[]{.99\linewidth}
      \includegraphics[width=1\linewidth]{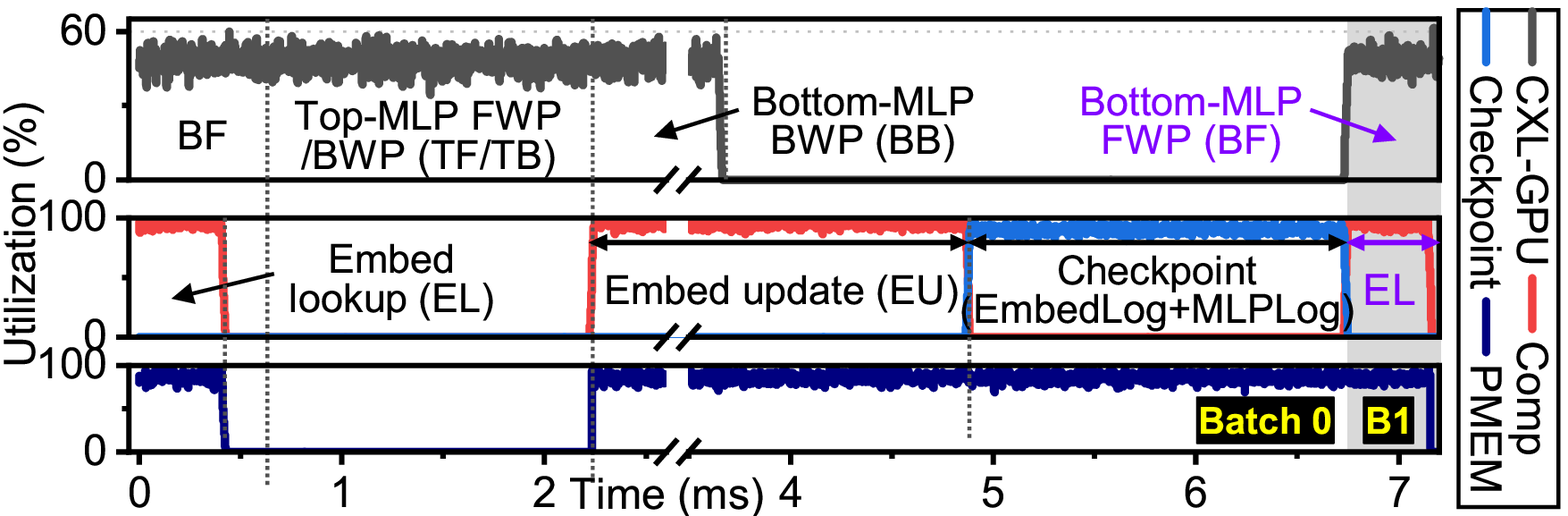}
      \caption{\texttt{CXL-D}.} \label{fig:eval_util1} \vspace{1pt}
    \end{subfigure}
    \begin{subfigure}[]{.99\linewidth}
        \includegraphics[width=1\linewidth]{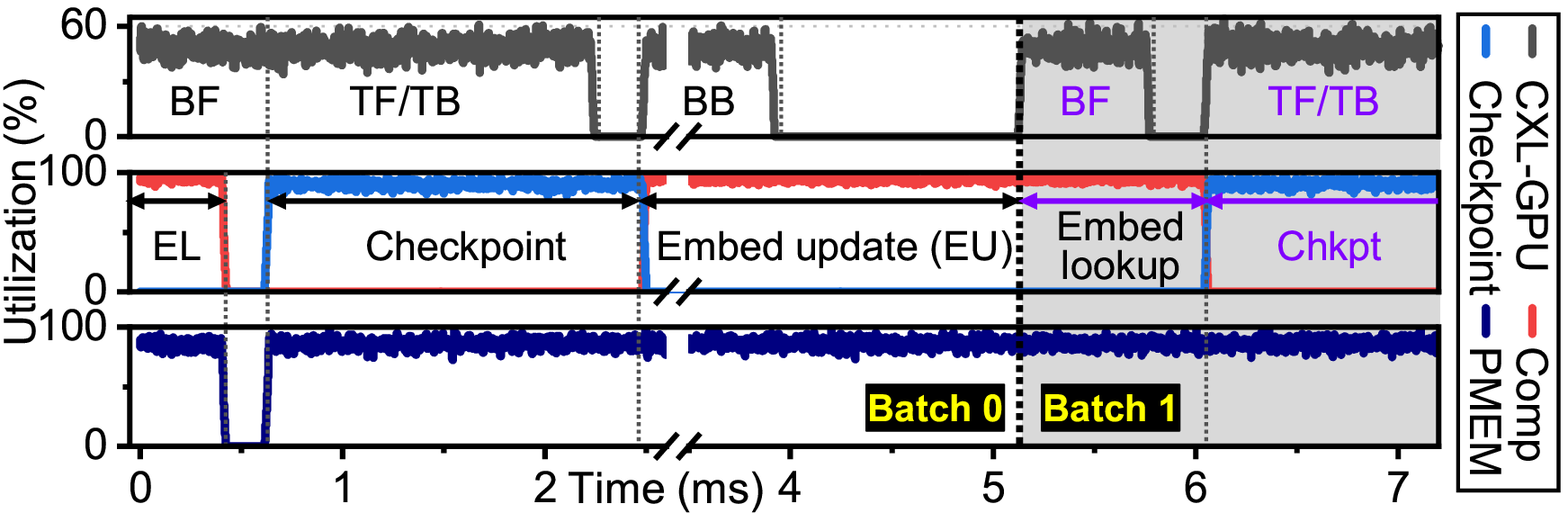}
        \caption{\texttt{CXL-B}.} \label{fig:eval_util2} \vspace{1pt}
    \end{subfigure}
    \begin{subfigure}[]{.99\linewidth}
        \includegraphics[width=1\linewidth]{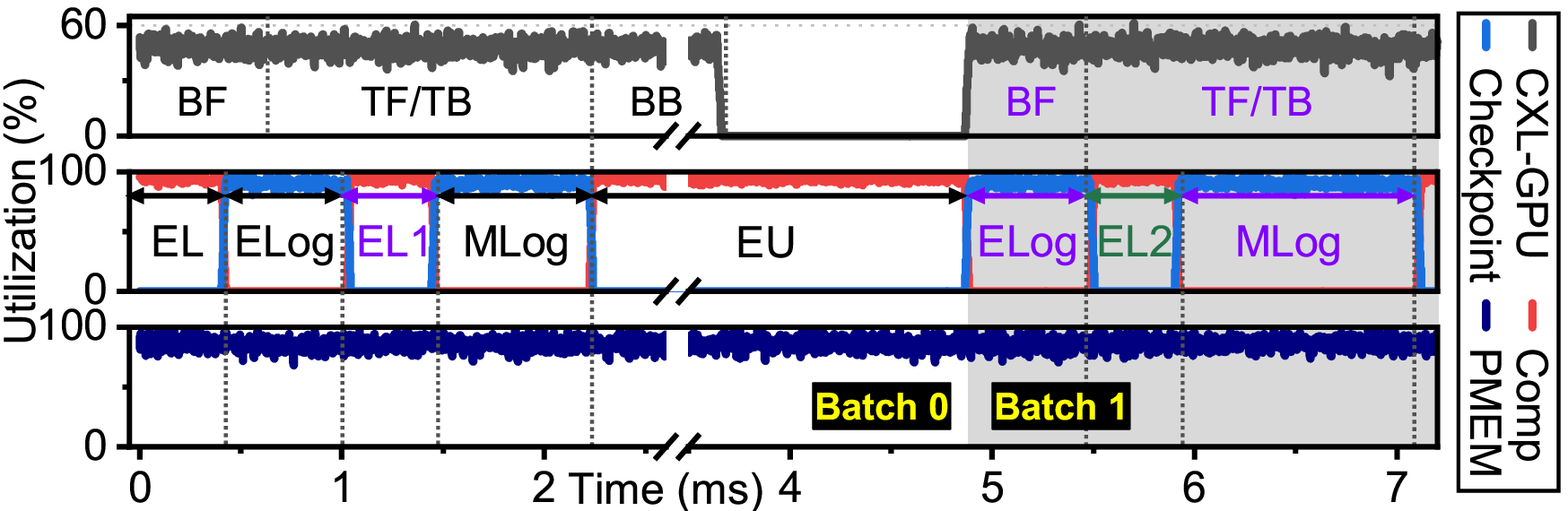}
        \caption{\texttt{CXL}.} \label{fig:eval_util3} \vspace{1pt}
    \end{subfigure}
    \caption{Utilization of hardware resources.} \label{fig:eval_util}
    \vspace{-10pt}
\end{figure}

We analyze the utilization of computing and memory resources, including CXL-GPU, CXL-MEM's computing logic, checkpointing logic, and PMEM.
We breakdown utilization timelines with RM training operations to understand the proposed contributions as shown in Figure \ref{fig:eval_util}.

Figure \ref{fig:eval_util1} shows training timeline of \texttt{CXL-D}.
During 0.6$ms$$\sim$2.2$ms$, CXL-MEM is idle while CXL-GPU performs feature interaction and top-MLP operations.
After CXL-MEM finishes the embedding update operation at 4.9$ms$, the checkpointing logic of CXL-MEM performs checkpointing in a redo log manner, and the next batch starts at 6.7$ms$.
Compared to \texttt{CXL-D}, \texttt{CXL-B} supports batch-aware checkpoints (cf. Figure \ref{fig:eval_util2}).
Therefore, when forward propagation of bottom-MLP is completed at 0.6$ms$, CXL-MEM's checkpointing logic starts to perform checkpointing.
Checkpointing overhead is observed during 2.2$ms$$\sim$2.5$ms$ since the checkpointing time is longer than the forward and backward propagation times of feature interaction and top-MLP operations.
Nevertheless, the training time is reduced by 1.6$ms$ compared to \texttt{CXL-D} because checkpointing is performed by fully utilizing the idle time of CXL-MEM.
Note that batch 1's embedding lookup time increases in \texttt{CXL-B} than \texttt{CXL-D}.
This is because PMEM's read-after-write performance is observed due to consequent embedding update (at batch 0) and embedding lookup (at batch 1) operations.

With the proposed training relaxation techniques, \texttt{CXL} not only maximizes computing and memory resource utilization but also improves training time (cf. Figure \ref{fig:eval_util3}).
Relaxed batch-aware checkpoint is observed during 1.4$ms$$\sim$2.2$ms$.
Since MLP logging starts from 1.4$ms$ and stops when the CXL-GPU's top-MLP operation is completed at 2.2$ms$, the checkpointing overhead is eliminated.
Relaxed embedding lookup is observed during 1$ms$$\sim$1.4$ms$.
By relaxing computational dependency to avoid read-after-write, embedding lookup time is shortened compared to \texttt{CXL-B}.

\subsection{Power Analysis}

\begin{figure}
	\centering
    \includegraphics[width=1\linewidth]{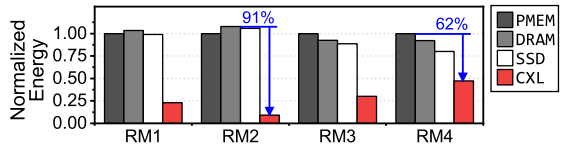}
	\vspace{-9pt}
	\caption{Energy analysis.} \label{fig:energy}
	\vspace{-12pt}
\end{figure}

In this subsection, we analyze energy consumption between \texttt{SSD}, \texttt{PMEM}, and \texttt{CXL}; the energy values of \texttt{SSD} and \texttt{CXL} are normalized to those of \texttt{PMEM} for better understanding.
For comparison with high-performance training systems, we also add \texttt{DRAM}, which is an ideal configuration where the entire embedding tables are loaded into DRAM.
As shown in Figure \ref{fig:energy}, our proposed \texttt{CXL} exhibits the lowest energy consumption across all RMs.
However, the energy consumption difference between \texttt{CXL} and other configurations varies greatly between RMs, based on the type of computation intensity.
For example, embedding-intensive RM2 shows the most significant energy savings, as much as 91\% compared to \texttt{DRAM}, whereas MLP-intensive RM4 only shows as much as 62\% compared to \texttt{PMEM}.
The reason why embedding-intensive RMs show more energy savings than MLP-intensive RMs is mainly related to how much CXL-MEM shortens the training time.

Although \texttt{DRAM} can train RMs faster than \texttt{SSD}, it consumes more energy than \texttt{SSD}.
This is because \texttt{DRAM} requires more memory modules to store the same size of embedding tables.
That's why the energy consumption of DRAM is much higher than that of PMEM during embedding table accesses.
However, when we compare the energy consumption of \texttt{DRAM} and \texttt{PMEM}, there are different trends between embedding-intensive models (RM1 and RM2) and MLP-intensive models (RM3 and RM4).
In RM1 and RM2, \texttt{DRAM} consumes more energy than \texttt{PMEM} due to DRAM's large energy consumption than PMEM.
On the other hand, the energy consumption of \texttt{PMEM} is higher than that of \texttt{DRAM} in RM3 and RM4.
This is because \texttt{PMEM} should log the large amount of bottom and top-MLP's parameters, whereas \texttt{DRAM} does not perform checkpointing.

\section{Related Work}
\label{sec:relatedwork}

\noindent \textbf{In-storage processing.}
Prior works proposed in-storage processing (ISP) to solve data movement overhead of SSD-based embedding table storage \cite{wilkening2021recssd, sun2022rm}.
While RecSSD \cite{wilkening2021recssd} considers embedding lookup as ISP, RM-SSD \cite{sun2022rm} offloads the entire recommendation system to the underlying storage device to eliminate the significant GPU-SSD synchronization overhead further.
In the sense that the CXL-MEM can alleviate data movement overhead, it is similar to the existing in-storage processing works.
However, \textsc{TrainingCXL} is differentiated from existing works as it guarantees a data persistency, and all data movement is done by hardware without software intervention.

\noindent \textbf{Near-memory processing.}
There are also DRAM-based near-memory processing schemes \cite{ke2020recnmp, kwon2019tensordimm} to achieve high-performance inference for recommendation systems.
However, they suffer from the limitation of memory expansion.
RecNMP \cite{ke2020recnmp} is implemented with the existing memory interface (e.g., DDR4), which limits the memory module expansion with the number of CPUs.
On the other hand, TensorDIMM \cite{kwon2019tensordimm} proposes a disaggregated memory node that can pool TensorDIMMs by leveraging NVSwitch.
However, there is a scalability limitation since NVSwitch cannot be employed in a multi-tiered manner.
\textsc{TrainingCXL} proposes a CXL-based memory disaggregation approach, which can easy to be expanded with the CXL 3.0's multi-level switching support.
In addition, thanks to the CXL-MEM's on-card computing capability, our work is differentiated from RecNMP and TensorDIMM, which exploits on-DIMM computing capability.
CXL-MEM can perform near-data processing without modification of commodity memory modules, and its computing capability is not limited by the memory media capacity.

\section{Conclusion}

\label{sec:conclusion}
We propose \textsc{TrainingCXL} that can efficiently process large-scale recommendation models in the disaggregated memory pool by integrating persistent memory and GPU into a single cache-coherent domain.
Our batch-aware checkpoint can effectively hide checkpointing overhead from the training process, and the relaxation of embedding lookup removes RAW conflict issues, thereby improving the training bandwidth.
Overall, our evaluation demonstrates that \textsc{TrainingCXL} achieves 5.2$\times$ training performance improvement while consuming 76\% lower energy compared to the state-of-the-art PMEM-based recommendation systems.

\section{Acknowledgement}
We appreciate anonymous reviewers and thanks for all the technical support of Panmnesia.
This work is protected by one or more patents.
Myoungsoo Jung is the corresponding author (\href{mailto:mj@camelab.org}{mj@camelab.org}).







\end{document}